\documentclass[epj]{svjour}

\usepackage{epsfig}
\usepackage{psfrag}
\usepackage{amsmath}
\usepackage{tabularx}

\renewcommand{\vec}[1]{\boldsymbol{\rm #1}}

\newcommand{\rhosand}{\rho_{\text{sand}}}
\newcommand{\rhoair}{\rho_{\text{air}}}
\newcommand{\rhoquartz}{\rho_{\text{quartz}}}

\newcommand{\veff}{v_{\text{eff}}}

\newcommand{\xlabel}[2]{
  \psfrag{#1}[t][]{#2}
}
\newcommand{\ylabel}[2]{
  \psfrag{#1}[c][t]{#2}
}

\begin{document}

\title{A Model of Barchan Dunes Including Lateral Shear Stress}

\author{V.~Schw\"ammle\inst{1} \and  H.J. Herrmann\inst{1,2}}
\institute{Institute for Computer Applications 1, University of Stuttgart,
Pfaffenwaldring 27, D-70569 Stuttgart, Germany \and
Departamento de F\'isica, Universidade Federal do Cear\'a,
60455-970 Fortaleza, Brazil}

\date{}


\abstract{
Barchan dunes are found where sand availability is low and wind direction quite constant.
The two dimensional shear stress of the wind field and the sand movement by saltation and
avalanches over a barchan dune are simulated. The resulting final shape is compared to the results 
of a model with a one dimensional shear stress. A characteristic 
edge at the center of the windward side is discovered which is also observed for big barchans. 
Diffusion effects reduce this effect for small dunes. 
} 

\maketitle

\section{Introduction}
\label{sec:barch_intro}

Barchan dunes are highly mobile dunes which constitute a considerable threat to  
infrastructure in arid regions with sand.
They move over roads, pipelines and cover even cities. The large time
scales involved  
considerably complicate reliable measurements. Nevertheless, over many decades measurements 
have been made of dunes all over the world 
\cite{Finkel59,Coursin64,Hastenrath67,Lettau69,Sarnthein74,%
Howard78,Jaekel80,Hastenrath87,Slattery90,Kocurek92,Wiggs96,%
Hesp98,Walker1998,Jimenez99,sauermann-etal:2000,unpub:SauermannAndrade2001}.
In order to learn more about the mechanism driving dune morphology and dynamics 
numerical models have been proposed \cite{Wippermann86,zeman-jensen:88,%
Fisher88,Stam97,NishimoriXX,boxel-arens-van_dijk:99,%
van_dijk-arens-boxel:99,herrmann-sauermann:2000,MomijiWarren2000,SauermannKroy2001,%
KroySauermann2002,unpub:SchwaemmleHerrmann2003}. The models revealed many interesting 
results and rised new questions which have to be answered. The inclusion of the lateral
component of the shear stress of the wind field over a barchan dune leads to the results 
presented in this paper. 

The following section will explain some aspects concerning barchan dunes. The model 
will be introduced in order to simulate barchan dunes. 
In the following section scaling relations calculated with the model will be presented and 
compared to measurements. The difference between the numerical results and 
real dune shapes  for small barchans will lead to an inclusion of diffusion in the 
saltation transport calculation. The last section will be about barchanoids. 

\section{Basics}
\label{sec:barch_bas}

The word {\em barchan} comes from the turkish language and means ``active dune''. It was
preserved in the scientific literature to name the isolated
crescent-shaped mobile dune. 
Less than 1\% of all dune sand on Earth is contained in barchan dunes. 
These 
dunes  exist mainly in areas where not very much sand is available and 
wind stays unidirectional.

\begin{figure}[htb]
  \begin{center}
    \includegraphics[width=0.4\textwidth]{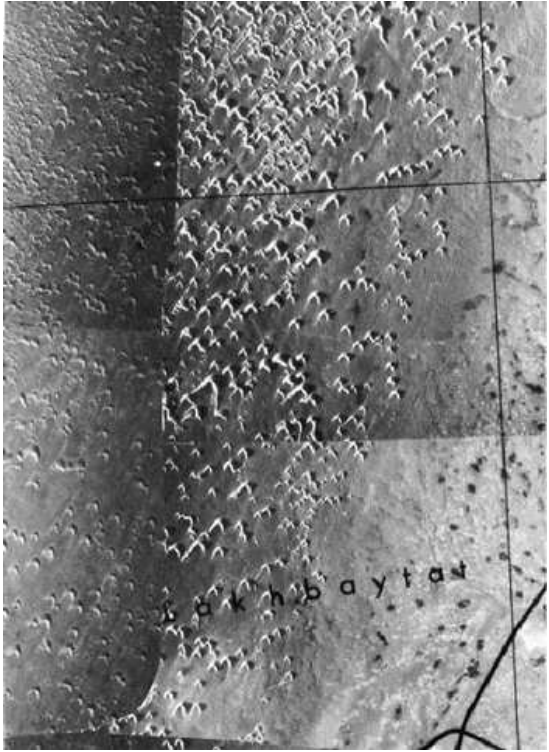}
\caption{Aerial photography of the dune field north of La\^ayoune, Morocco. The dune heights 
       are of some meters. The wind is blowing from NNE to SSW (photo
       taken from Sauermann (2001)).}
    \label{fig_barch_Morocco}
\end{center}
\end{figure}

\begin{figure}[htb]
  \begin{center}
    \includegraphics[width=0.4\textwidth]{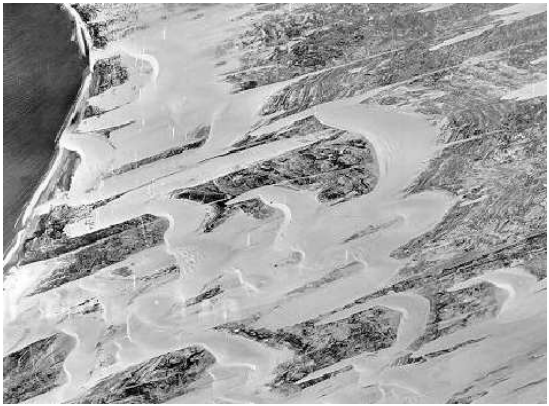}
\caption{Aerial photography of the dune field of Jericoacoara near Fortaleza, Brazil. 
         Some dunes are over 50 meters high.
         Photo taken from Jimenez  et al (1999).}
    \label{fig_barch_jeri}
\end{center}
\end{figure}

The size of barchans varies from heights of some meters 
(Figure~\ref{fig_barch_Morocco}) to over 50 m (Figure~\ref{fig_barch_jeri}). 
Barchan dunes are not exactly shape invariant. There is a minimal height of 1-2 meters 
below which it looses its sand and no stable shape is reached. Dune shapes
seem to be controlled by the saturation length of the saltation transport on the windward
side \cite{SauermannKroy2001,Hersen2002}. This  is the distance
needed to reach the saturated sand flux over a sandy surface.
The saturation length, not to be confused with the  
saltation length (mean length of a grain trajectory in air), has a complex dependency
on the air shear stress. Zero flux over a ground without 
sand bed needs a transient length  to reeach a saturated flux  at the place where the 
surface begins to be covered with sand.
 
Small stable barchan dunes have a short slip face and
the crest does not coincide with the brink. Whereas the sand is trapped completely in the 
slip face, the lack of a slip face at the horns allows the sand grains
to leave the dune there.
Thus a barchan dune grows if sand influx is larger than outflux and vice versa. 
%

\section{The model}
\label{sec:model}

The model described here can be seen as a minimal model including the main processes of 
dune morphology. 
As predecessors of the model described here the works of \cite{SauermannPhD2001} 
and \cite{unpub:SchwaemmleHerrmann2003} revealed interesting new insights 
into dynamics and 
formation of dunes. The model of \cite{unpub:SchwaemmleHerrmann2003}, modelling 
transverse dunes, will be used here to obtain the steady state of barchan dunes.
The simulations are carried out with a completely unidirectional  and constant wind source.
In this article the model is used to simulate isolated single barchan dunes.
In every iteration the horizontal shear stress $\vec{\tau}$ of the wind, the saltation flux 
$\vec q$ and the flux due to avalanches are calculated. The time scale of these processes 
is much shorter than the 
time scale of changes in the dune surface so that they are treated to be instantaneous. 
In the following the different steps at every iteration are explained. 

\paragraph{The air shear stress $\tau$ at the ground:}

We use the well known logarithmic velocity profile of the atmospheric boundary layer as a 
basis of our shear stress calculation,
\begin{equation}
  \label{eq:v_z_log}
  v(z) = \frac{u_*}{\kappa} \ln \frac{z}{z_0},
\end{equation}
where $v(z)$ is the horizontal wind velocity, $z_0$ the roughness length, giving a measure of the
roughness of the surface, and $u_* = \sqrt{\tau / \rho}$ is called shear velocity. Although it 
has the dimension of a velocity the shear velocity $u_*$  is used as a measure for the 
shear stress. The shear stress perturbation over a single dune or 
over a dune field given by the height $h(x,y)$ is calculated using
the algorithm of \cite{weng-etal:91}. The $\tau_x$-component points in wind direction and the 
$\tau_y$-component denotes the lateral direction perpendicular to the wind. The calculation is made in Fourier space, 
$k_x$ and $k_y$ are the wave numbers,
\begin{equation}
  \label{eq:tau_x}
  \hat\tau_x(k_x,k_y) = 
    \frac{h(k_x,k_y) k_x^2}{|k|} 
  \frac{2}{U^2(l)} \cdot  \nonumber
\end{equation}
\begin{equation}
  \left( 1 +   
   \frac{2 \ln L|k_x| + 4 \gamma + 1 +
      i \, \text{sign}(k_x) \pi}{\ln l/z_0} \right),
\end{equation}
and
\begin{equation}
  \label{eq:tau_y}
  \hat\tau_y(k_x,k_y) = \frac{h(k_x,k_y) k_x k_y}{|k|}
  \frac{2}{U^2(l)},
\end{equation}
where  $|k| = \sqrt{k_x^2 + k_y^2}$ and $\gamma=0.577216$ 
(Euler's constant). $U(l)$ is the normalized velocity of the undisturbed logarithmic 
profile at the height of the inner region $l$ \cite{SauermannPhD2001} defined in 
\cite{weng-etal:91}. The values of the roughness length $z_0$ and the so called 
characteristic length $L$ were adjusted to the data of 
\cite{sauermann-etal:2000} supposing we have an averaged shear velocity of $u_*=0.5$ ms$^{-1}$
in the region where the measurements were carried out.
Thus $z_0$ is set to $0.0025$ m and $L$ to $10$ m. 

Equations~(\ref{eq:tau_x}) and (\ref{eq:tau_y}) are calculated in Fourier space 
and have to be multiplied with the logarithmic 
velocity profile of Equation~(\ref{eq:v_z_log}) in real space in order to obtain the 
total shear stress. The surface is 
assumed to be instantaneously rigid and the effect of 
sediment transport is incorporated in the roughness length $z_0$. 
For slices in wind direction the separation streamlines in the lee zone
of the dunes are fitted by a polynomial of third order attaching to the brink continuously. 
The length of the separation streamlines is determined by allowing a maximum 
slope of $14^o$ \cite{SauermannPhD2001}. The separation bubble guarantees 
a smooth surface for the wind field calculation and the shear stress in the area inside the 
separation bubble is set equal to 
zero. Problems can arise due to numerical fluctuations in the value of the windward
slope at the brink where the separation bubble begins and its influence on the calculation 
of the separation streamline for each 
slice. To get rid of this numerical error the surface is Fourier-filtered by cutting the 
small frequencies. 

\paragraph{The saltation flux $q$:}

The product of the sand density and the sand velocity leads to the sand flux over 
the surface, $\vec q(x,y)=\vec u(x,y) \rho(x,y)$. The saltation flux is calculated from 
 mass conservation,
\begin{equation}
  \frac{\partial \rho(x,y,t)}{\partial t}
  + \nabla \rho(x,y,t) \vec u(x,y,t) \nonumber
\end{equation}
\begin{equation}
  + C_{diff} \vec{\Delta} \rho(x,y,t)
  = \Gamma(x,y,t),
  \label{eq:rho_s_0}
\end{equation}
where $\rho$ is the density and 
$\vec u$ the velocity of the sand grains in the saltation layer. These variables are integrated
over the vertical component. $C_{diff}$ denotes a diffusion constant which
will be the crucial parameter in
Section~\ref{sec:barch_effect_diff}. In the other chapters 
$C_{diff}$ is set to zero.
Grains are entrained by the air and deposited on the sand bed which is expressed by the 
exchange term $\Gamma$.
The velocity of the sand grains is calculated from momentum conservation,
\begin{equation}
  \frac{\partial \vec u(x,y,t)}{\partial t} 
  + \left( \vec u(x,y,t) \nabla \right) \vec u(x,y,t) 
  = \nonumber
\end{equation}
\begin{equation}
\frac{1}{\rho(x,y,t)} \left( \vec f_{drag}(x,y,t) + \vec f_{bed}(x,y,t) + 
        \vec f_g(x,y) \right).
  \label{eq:u_s_0}
\end{equation}
The drag force $\vec f_{drag}$ denotes the interaction between air flow and sand grains, 
$\vec f_{bed}$ is the decceleration of the grains by the collition  with the sand bed and
$\vec f_g$ is the gravity force which drives the grains into the direction of the 
 steepest gradient of the
height profile. 
The time to reach the steady state of sand flux over a new surface
is several orders of magnitude shorter than the time scale of the surface evolution.
Hence, the steady state is assumed to be reached instantaneously. 
The  characteristic length scale to reach a saturated saltation layer influences the calculation by breaking the
scale invariance of dunes and by determining the minimal size of a barchan dune 
\cite{SauermannKroy2001}.

A calculation of the saltation transport by the well known flux relations 
\cite{Bagnold41,Lettau78,Sorensen91} would restrict the model to saturated sand flux which
is not the case  observed at the foot of the windward side of a
barchan dune due to the reduced 
sand supply. We simplified the closed model of 
\cite{SauermannKroy2001}, resulting from Equation~(\ref{eq:rho_s_0}) and (\ref{eq:u_s_0}),
by neglecting the time dependent terms and the convective term 
of the grain velocity $\vec u(x,y)$. This
yields Equations~(\ref{eq:3d_rho}) and (\ref{eq:3d_u}) where $\rho$ and 
$\vec u$ are determined from the  shear stress obtained before  and the gradient of the 
actual surface, 
\begin{equation}
  \label{eq:3d_rho}
  \text{div} \, (\rho \, \vec u) + C_{diff} \Delta \rho 
  = \frac{1}{T_s} \rho \left( 1 - \frac{\rho}{\rho_s} \right) \; 
  \begin{cases}
    \Theta(h) & \rho < \rho_s\\
    1         & \rho \ge \rho_s
  \end{cases}
  ,
\end{equation}
with
\begin{equation}
  \label{eq:rho_s_tau}
  \rho_s = \frac{2 \alpha}{g} \left( |\vec \tau| - \tau_t \right) \quad \quad
  T_s = \frac{2 \alpha | \vec u|}{g} \, \frac{\tau_t}{\gamma (|\vec \tau| - \tau
_t)}.
\end{equation}
and
\begin{equation}
  \label{eq:3d_u}
   \frac{3}{4} \, C_d \frac{\rho_{\text{air}}}{\rho_{\text{quartz}}} d^{-1} \, (
\vec \veff - \vec u)|\vec \veff - \vec u|
    - \frac{g}{2 \alpha} \frac{\vec u}{|\vec u|}
    - g \, \vec \nabla \, h
    = 0,
\end{equation}
where $v_{\text{eff}}$ is the velocity of the grains in the saturated state,\\
\begin{equation}
 \vec v_{\text{eff}} =   \frac{2 \vec u_*}{\kappa |\vec u_*|} \cdot \nonumber 
\end{equation}
\begin{equation}
  \label{eq:3d:v_eff_of_tau_g0}
\left( \sqrt{\frac{z_1}{z_m} u_*^2 + 
       \left( 1 - \frac{z_1}{z_m} \right)  
      \, u_{*t}^2} 
     + \left( 
     \ln{\frac{z_1}{z_0}} -2 \right) \, \frac{u_{*t}}{\kappa} 
  \right),
\end{equation}
and
\begin{equation}
  \label{eq:ustar_s}
  u_* = \sqrt{\tau / \rhoair}
\end{equation}
In Equation~(\ref{eq:3d_u}) the force terms are arranged in the same order as in 
Equation~(\ref{eq:u_s_0}).
The constants and model parameters have been taken from \cite{SauermannKroy2001} and are 
summarized here:  $g=9.81\,$m$\,$s$^{-2}$,$\kappa=0.4$,
$\rhoair=1.225\,$kg$\,$m$^{-3}$,
$\rhoquartz=2650\,$kg$\,$m$^{-3}$, $z_m=0.04\,$m,
$z_0=2.5~10^{-5}\,$m, $D=d=250\,\mu$m, $C_d=3$,
$u_{*t}=0.28\,$m$\,$s$^{-1}$, $\gamma=0.4$, $\alpha=0.35$ and $z_1=0.005\,$m.

\paragraph{Avalanches:}

Surfaces with slopes which exceed the maximal stable angle of a sand surface, the called 
{\em angle of repose} $\Theta \approx 34^o$, produce avalanches which slide down in the 
direction of the steepest descent. The unstable surface relaxes to a somewhat smaller 
angle. For the study of dune formation two global properties are of interest. 
These are the sand transport downhill due to gravity and the maintenance of the angle of
repose. To determine the new surface after the relaxation by avalanches  the model 
proposed by \cite{Bouchaud94} is used. For the equations see 
\cite{unpub:SchwaemmleHerrmann2003}. 
%
Like in the calculation of the sand flux the steady state of the avalanche model
is assumed to be reached instantaneously. Hence, we can neglect the time dependent terms.
In the dune model a certain amount of sand 
is transported over the brink to the slip face and in every iteration
the sand excess is relaxed over the slip face by this avalanche model determining the steady state. 
\paragraph{The time evolution of the surface}

The calculation of the sand flux over a not stationary dune surface leads to 
changes by erosion and deposition of sand grains. The change of the surface profile can 
be calculated using the conservation of mass,
\begin{equation}
  \label{eq:masscons_h}
  \frac{\partial h}{\partial t} = - \frac{1}{\rhosand} \nabla{\vec q}.
\end{equation}
Finally, it is noted that Equation~(\ref{eq:masscons_h}) is the only remaining time dependent
equation and thus defines the characteristic time scale of the model which is normally
between 3--5 hours for every iteration.

\paragraph{The initial surface and boundary conditions:}

The genesis of a dune is still not known very well. Hence, 
the simulations must start with an essentially arbitrary initial surface. We can then observe
how the system reaches a final steady state. A steady state means that the dune shape does not undergo temporal
changes anymore.

As initial surface we take a flat bed of solid ground with a Gaussian hill of sand on it.
The maximum slope of the Gaussian is restricted in order to stay within the approximations
of the shear stress calculation.

We use quasi--periodic boundary conditions, i.e. the sand outflux at the outlet of the
simulation field is integrated and the same amount of sand enters the field inlet as a constant
influx.  Neglecting the amount of sand 
leaving the simulation area by the lateral boundary, the sand volume is conserved.

\section{Scaling laws}
\label{sec:barch_big_barchan}

In this section the morphologic relationships between height $h$, width $w$, length $l$ and 
velocity $v_d$
of barchan dunes resulting of the simulations are presented. The shapes for different 
dune sizes are compared. Therefore calculations of dunes which have different sizes were 
performed for the shear velocities $u_*=0.4$ ms$^{-1}$, $u_*=0.45$ ms$^{-1}$ and 
$u_*=0.5$ ms$^{-1}$. Single isolated barchan dunes are modeled using a quasi-periodic 
boundary until the final steady shape is reached.

\begin{figure}[htb]
  \center
    \includegraphics[angle=270,width=0.45\textwidth]{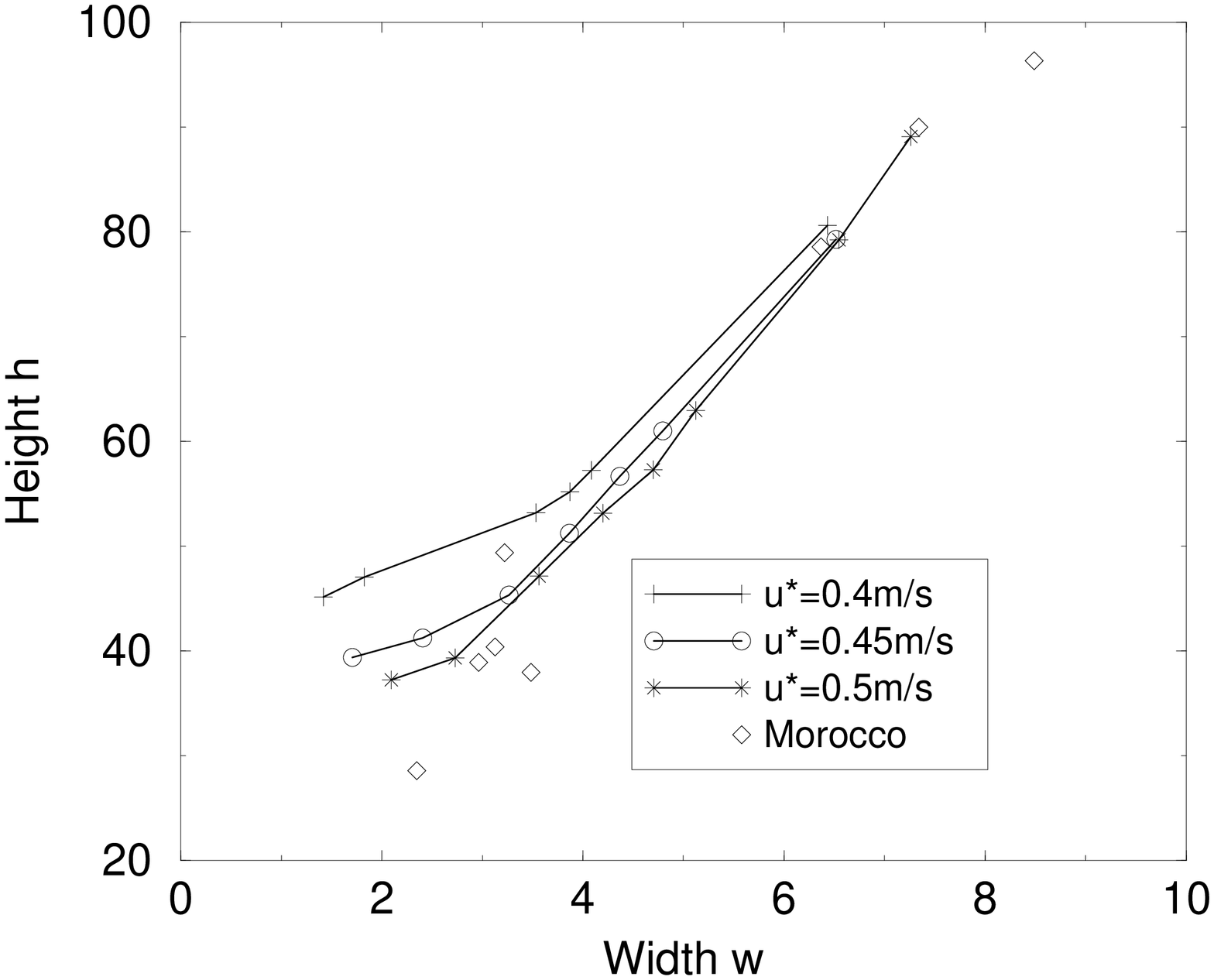}
\caption{Height width relation of dunes in Morocco (diamonds) and the results of numerical 
         calculations for different shear velocities.}
    \label{fig_barch_hw}
\end{figure}

\begin{figure}[htb]
  \center
    \includegraphics[angle=270,width=0.45\textwidth]{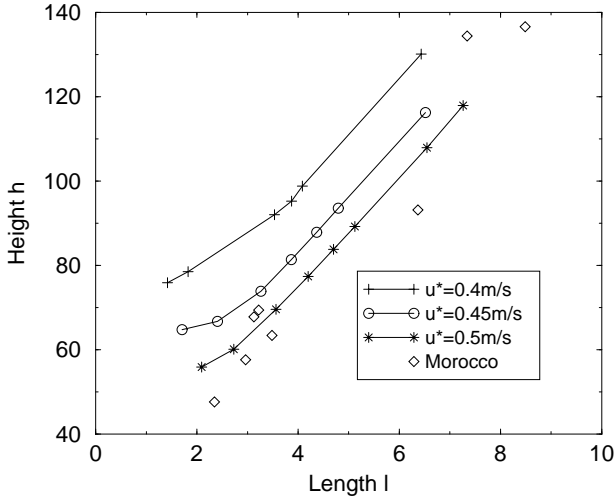}
\caption{Height length relation of dunes in Morocco (diamonds) and the results of numerical 
         calculations for different shear velocities.}
    \label{fig_barch_hl}
\end{figure}

\paragraph{Height, width and length relationships:}
Linear relationships between height, width and length were observed by
\cite{Finkel59} and \cite{Hesp98} for barchans in southern Peru and by 
\cite{herrmann-sauermann:2000} in Morocco.
Figures~\ref{fig_barch_hw} and \ref{fig_barch_hl} depict the height width and the height
length relationships for different shear 
velocities, respectively. 

A linear relationship is obtained for dunes larger than 2--3 meters. 
The linearity is valid only for length 
scales much larger than the saturation length ($l \gg l_s$). The functional 
dependence of the saturation length on the shear velocity predicts larger width and length for a dune of the same height,
for decreasing shear velocities. This is in agreement with \cite{SauermannKroy2001} who 
found an increase of $l_s$ at the shear velocity threshold.

The asymptotic slopes of the linear relationships are not affected by different shear velocities. 
The results are compared to the data of \cite{herrmann-sauermann:2000}
(Figures~\ref{fig_barch_hw} and \ref{fig_barch_hl}). 
The data of the barchans of Morocco fit best to the simulation results for a shear
velocity $u_*=0.5$ ms$^{-1}$.  Table~\ref{tab:H_W} compares the slope $a_W$ 
and the axis intercept $b_W$ of the linear relationships to the data of 
\cite{herrmann-sauermann:2000}, \cite{Hastenrath67} and 
\cite{Finkel59}.  
\begin{table}[bp]
    \caption{Height and width relationship.}
    \vspace{0.2cm}
    \begin{tabularx}{\linewidth}{p{0.4\linewidth}XX}
      \hline
                &  $a_W$ & $b_W$\\
      \hline 
        Finkel     & 10.3 & 4.0 \\
        Hastenrath &  8.2 & 9.5 \\
        Herrmann   & 11.1 & 5.6 \\
      \hline
        Simulations: & & \\
        $u_*=0.4$ ms$^{-1}$ & 9.9 & 16.6 \\
        $u_*=0.45$ ms$^{-1}$ & 10.6 & 10.4 \\
        $u_*=0.5$ ms$^{-1}$ & 10.5 & 8.5 \\
      \hline
    \end{tabularx}
    \label{tab:H_W}
\end{table}

\paragraph{Dune velocity:}

\begin{figure}[tb]
  \center
    \includegraphics[width=0.45\textwidth]{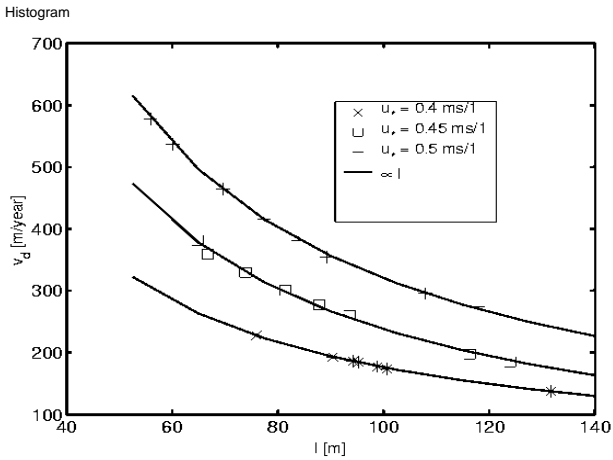}
\caption{The velocity $v_d$ of barchans fits very well to their reciprocal length. Note: The
         velocity is assumed to be constant for 365 days of wind per year and is thus smaller in real
         conditions.}
         
    \label{fig_barch_vdune}
\end{figure}

\begin{figure}[htb]
  \center
    \xlabel{normalized width y}{normalized width $\tilde{y}$}
    \ylabel{normalized height z}{normalized height $\tilde{z}$}
    \includegraphics[width=0.45\textwidth]{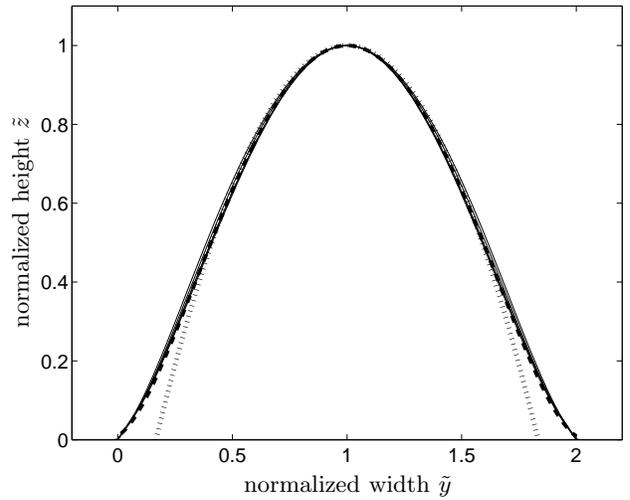}
\caption{Normalized transversal profiles of the dunes. A fit with $\cosh^2(x)$ is given by
         the dashed line and fits much better than  a parabola (dotted line). 
         The shear velocity is $u_*=0.5$ms$^{-1}$.}
    \label{fig_barch_bigcy2}
\end{figure}

\begin{figure}[htb]
  \center
    \xlabel{normalized length x}{normalized length $\tilde{x}$}
    \ylabel{normalized height z}{normalized height $\tilde{z}$}
    \includegraphics[width=0.45\textwidth]{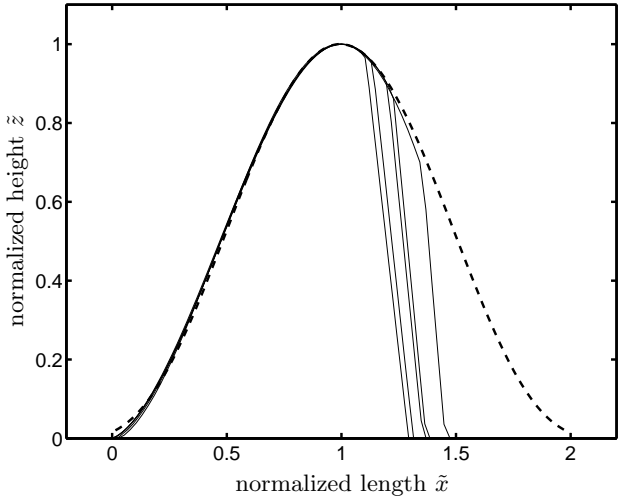}
\caption{Normalized longitudinal profiles of the dunes. A fit with $\cosh^4(x)$ (dashed line)
         reproduces quite well the windward side. The shear velocity is $u_*=0.5$ms$^{-1}$.}
    \label{fig_barch_bigcx}
\end{figure}

\begin{figure}[htb]
  \center
    \includegraphics[width=0.45\textwidth]{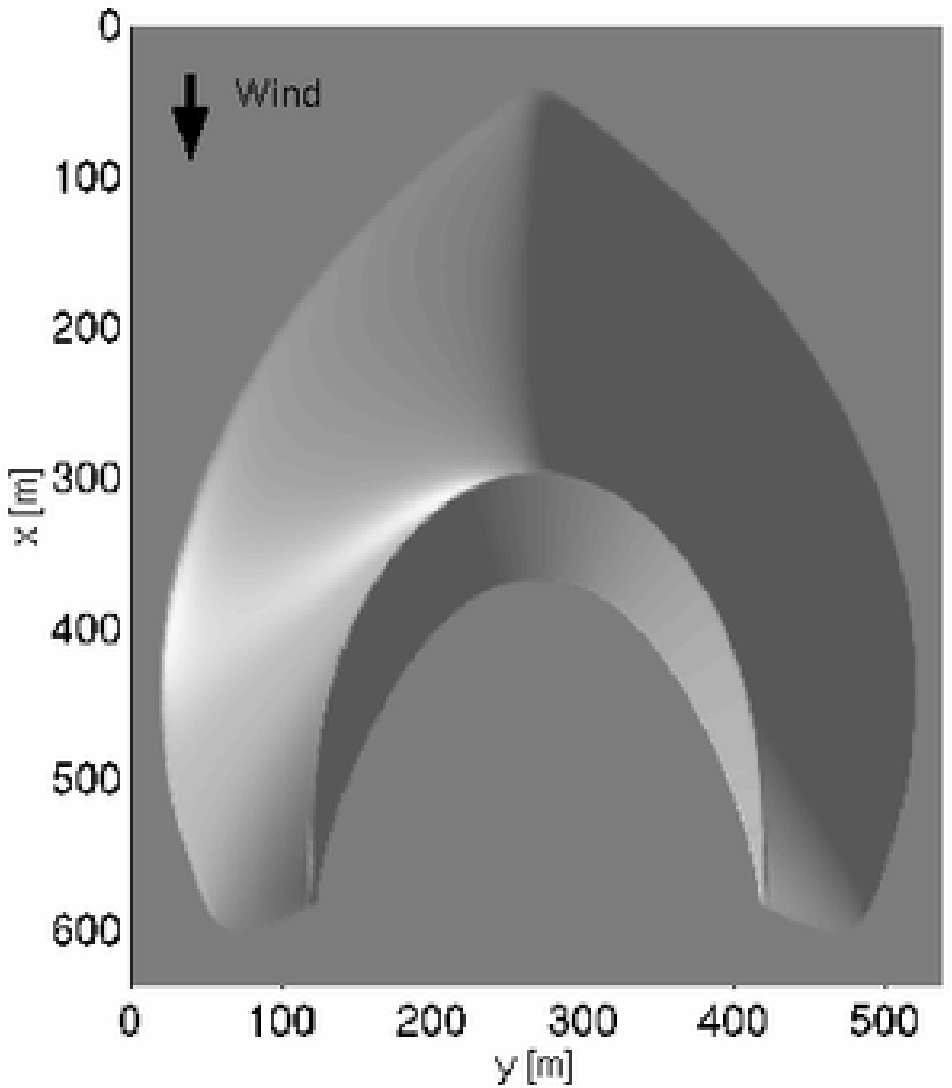}
    \includegraphics[width=0.45\textwidth]{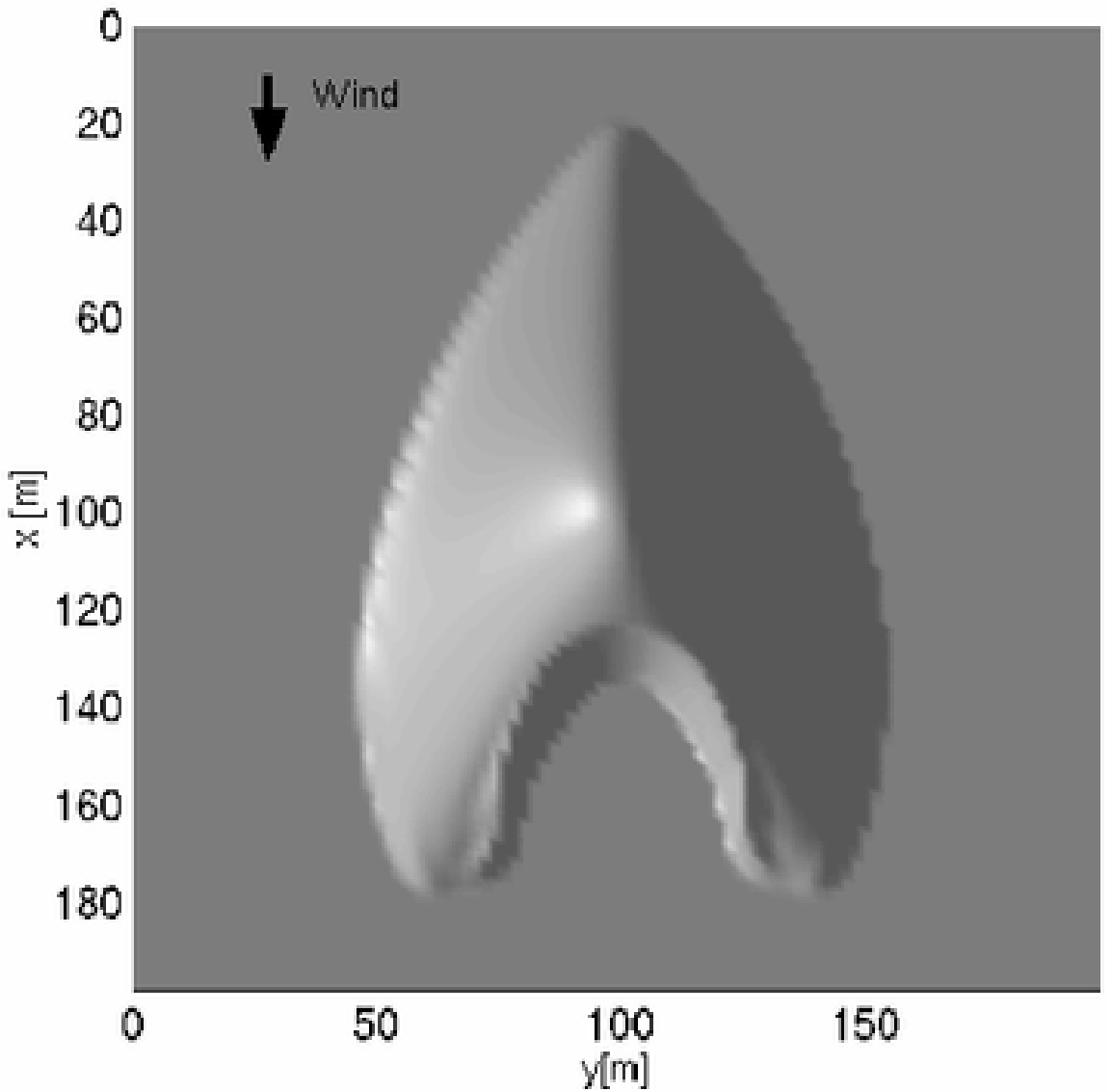}
\caption{On top the surface of a $45$ m high barchan dune is depicted. The dune in the
         bottom part has a height of $5$ m. There is no simple shape invariance. 
         The shear velocity is $u_*=0.5$ms$^{-1}$ for both simulations.}
    \label{fig_barch_big1surf2}
\end{figure}

According to \cite{SauermannPhD2001} the dune velocity decreases 
inversely proportional to the length of the envelope of the surface formed by the height profile
and the separation bubble. The separation bubble of the barchans in the dune model described
here fills almost exactly the region between the horns so that the
length $l$ of the dune can be used in order to evaluate the simulation results. Thus 
Bagnold's law, the reciprocal proportionality of the dune velocity $v_d$ 
to the height $h$, has to be modified, to 
\begin{equation}
  v_d = \frac{\Phi_{dune}}{l}
\label{eq:v_dune_b}
\end{equation}
where $\Phi_{dune}$ is in principle dependent on the  shear velocity.
The relation between dune velocity and height revealed the same
deviations from Bagnold's law for small dunes as in the case of the height length 
relationship (Figure~\ref{fig_barch_hl}). Figure~\ref{fig_barch_vdune} depicts the dune 
velocity $v_d$ versus dune length 
$l$. The surprising result is that the dune velocity does not differ from 
Equation~\ref{eq:v_dune_b} even for small dune sizes and so the saturation length seems to 
have no influence on the dune velocity. 
%
%
\begin{figure}[htb]
  \center
    \xlabel{normalized width y}{normalized width $\tilde{y}$}
    \ylabel{normalized height z}{normalized height $\tilde{z}$}
    \includegraphics[width=0.45\textwidth]{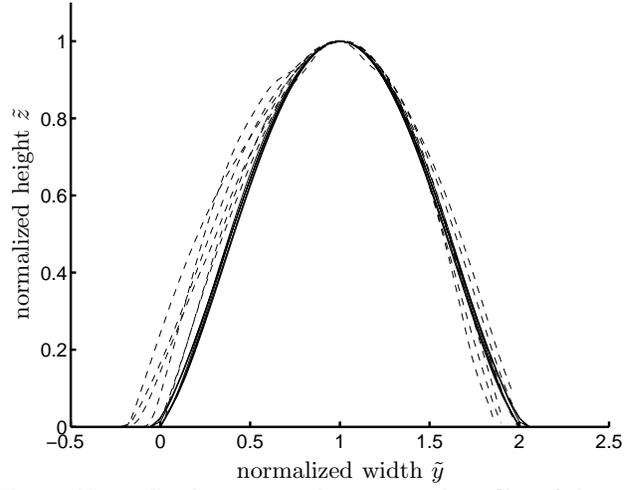}
\caption{Normalized superposed transversal profiles of the simulations (solid lines) and normalized
         transversal profiles of the dunes measured in Morocco from Herrmann and Sauermann (2000)        (dashed lines).
         The shear velocity in the simulations is $u_*=0.5$ms$^{-1}$.}
    \label{fig_barch_bigcy}
\end{figure}
\paragraph{The shape:}
Barchans are not really shape invariant due to the important role of the saturation length of the 
saltation transport. Nevertheless the shapes of barchans of
different sizes are compared in order to obtain more information about the deviation from
shape invariance. \cite{herrmann-sauermann:2000} fitted the normalized transversal 
profile of the measured dunes in Morocco with a parabola.
\cite{SauermannPhD2001} 
found a good agreement of the normalized transversal profile of his
numerical calculations  
(neglecting the lateral shear stress) to a 
parabola. Following his work the axes are rescaled and dimensionless variables are introduced,
\begin{equation}
  \label{eq:rescale}
  \tilde{x} = \frac{1}{l} x \quad
  \tilde{y} = \frac{1}{w} y \quad
  \tilde{z} = \frac{1}{h} z .
\end{equation}
Figure~\ref{fig_barch_bigcy} shows our normalized transversal profiles compared with the field
data of \cite{SauermannPhD2001}. In Figure~\ref{fig_barch_bigcy2} we compare the results
of the normalized transversal profiles with a fit to a parabola.
The upper part of the profile fits quite well whereas the lower part is far away from a
parabola. We searched for a better fit 
with another function. Figures~\ref{fig_barch_bigcy2} and \ref{fig_barch_bigcx} indicate that
the powers of a $\cosh$--function fit quite well, even the longitudinal profile. 
The inclusion
of lateral shear stress in the model seems to lead  qualitatively at
least for the transversal  cuts to a change from a parabolic 
to a $\cosh^2$--profile. 
In Figure~\ref{fig_barch_bigcx} the slope at the brink at the windward side
decreases for larger dunes. 
Finally the shapes of a $45$ and a $5$ meter high dune are depicted in 
Figure~\ref{fig_barch_big1surf2}. The slip face of larger dunes cuts a larger piece from
the dune body.

In figure~\ref{fig_barch_big1surf2} we can see a rather sharp edge in  the shape at the center of the 
windward side which is 
not observed in the results of the dune model of \cite{SauermannPhD2001}. 
The inclusion of the lateral component in the shear stress calculation
seems  responsible for this 
difference. This edge has not been observed for small dunes, for example in Morocco. 
But the large dunes in Figure~\ref{fig_barch_jeri} show a very similar
edge. Exact measurements of the shape of a large barchan 
would reveal more information and hopefully validate the model results. We assume that 
the absence of the edge for small dunes can be explained by diffusion-like smoothening effects occuring  during the 
saltation transport. Diffusion which in the simulations depicted so far has not been
considered acts on small scales and for larger scales the influence of diffusion on the dune
shape should be negligible. In the next section the shapes obtained
from  the model are studied for 
different diffusion constants.
\begin{figure}[htb]
  \center
    \includegraphics[width=0.45\textwidth]{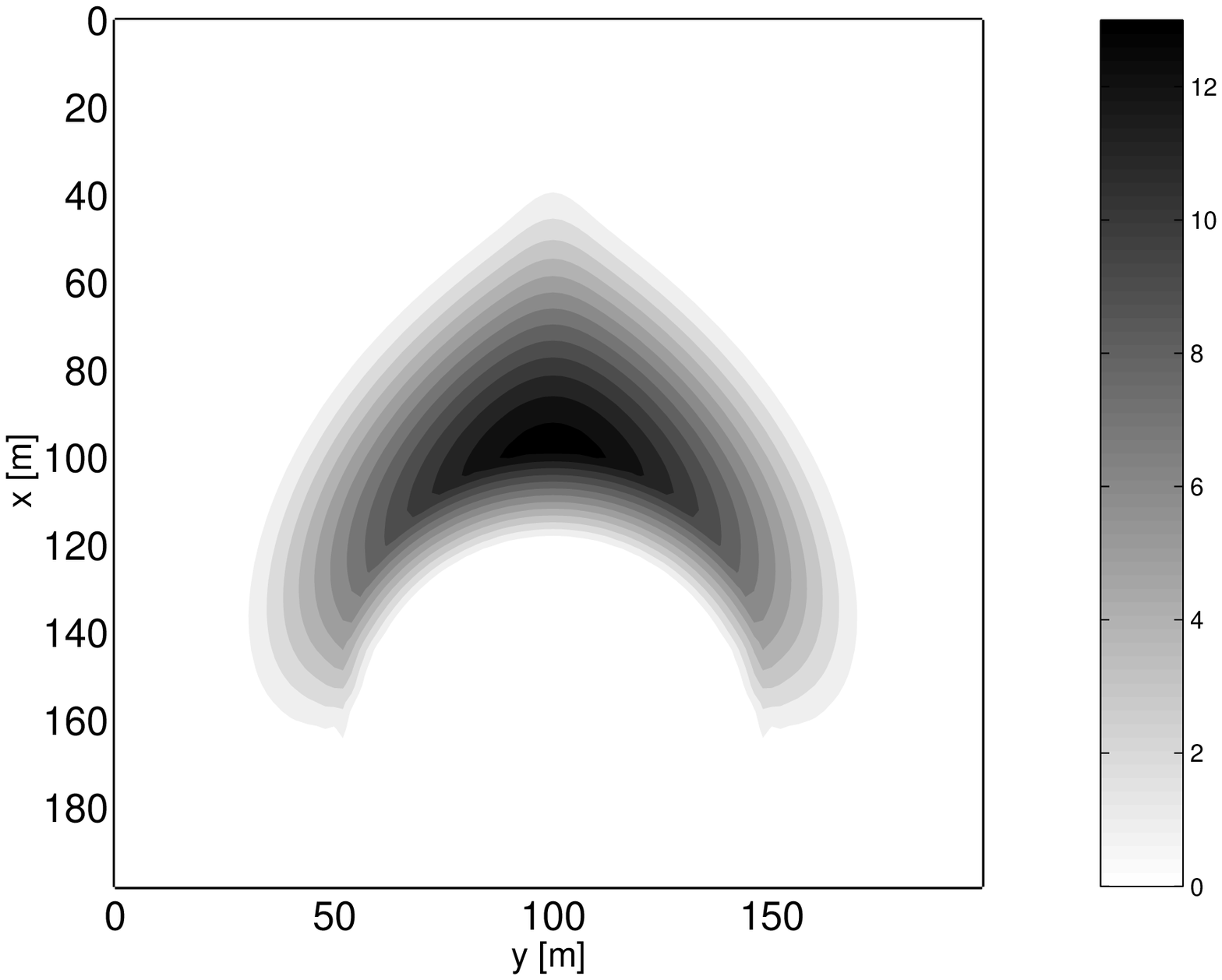}
    \includegraphics[width=0.45\textwidth]{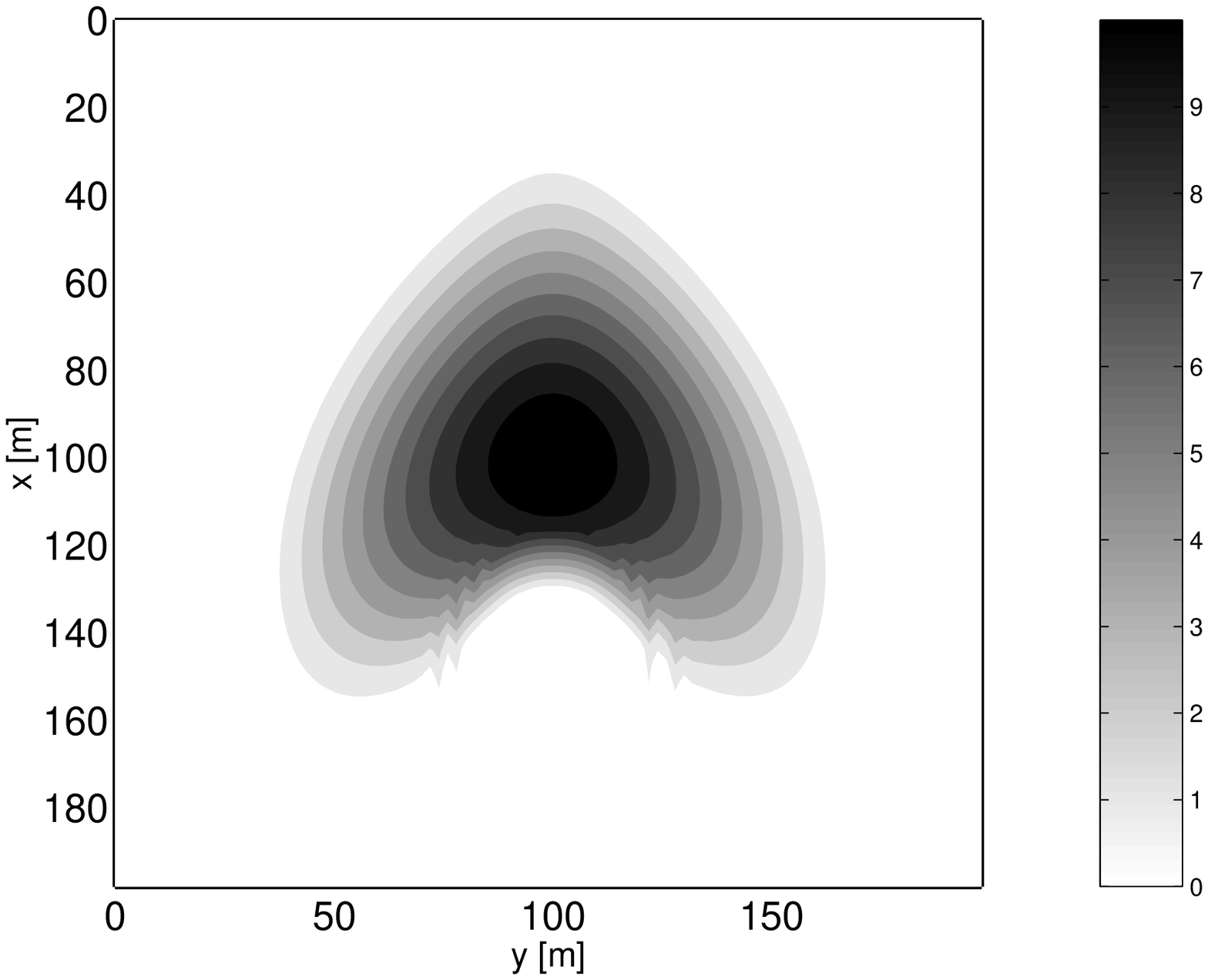}
\caption{Comparing the topography of two barchan dunes of equal volume
in steady state calculated from 
         the same initial surface for 
         a diffusion constant of $C_{diff} = 0.0$ (top) and $C_{diff} = 4.0$ m$^2$s$^{-1}$ 
         (bottom). The shear velocity is $u_*=0.5$ms$^{-1}$.}
    \label{fig_barch_diffh}
\end{figure}
\begin{figure}[htb]
  \center
    \includegraphics[width=0.45\textwidth]{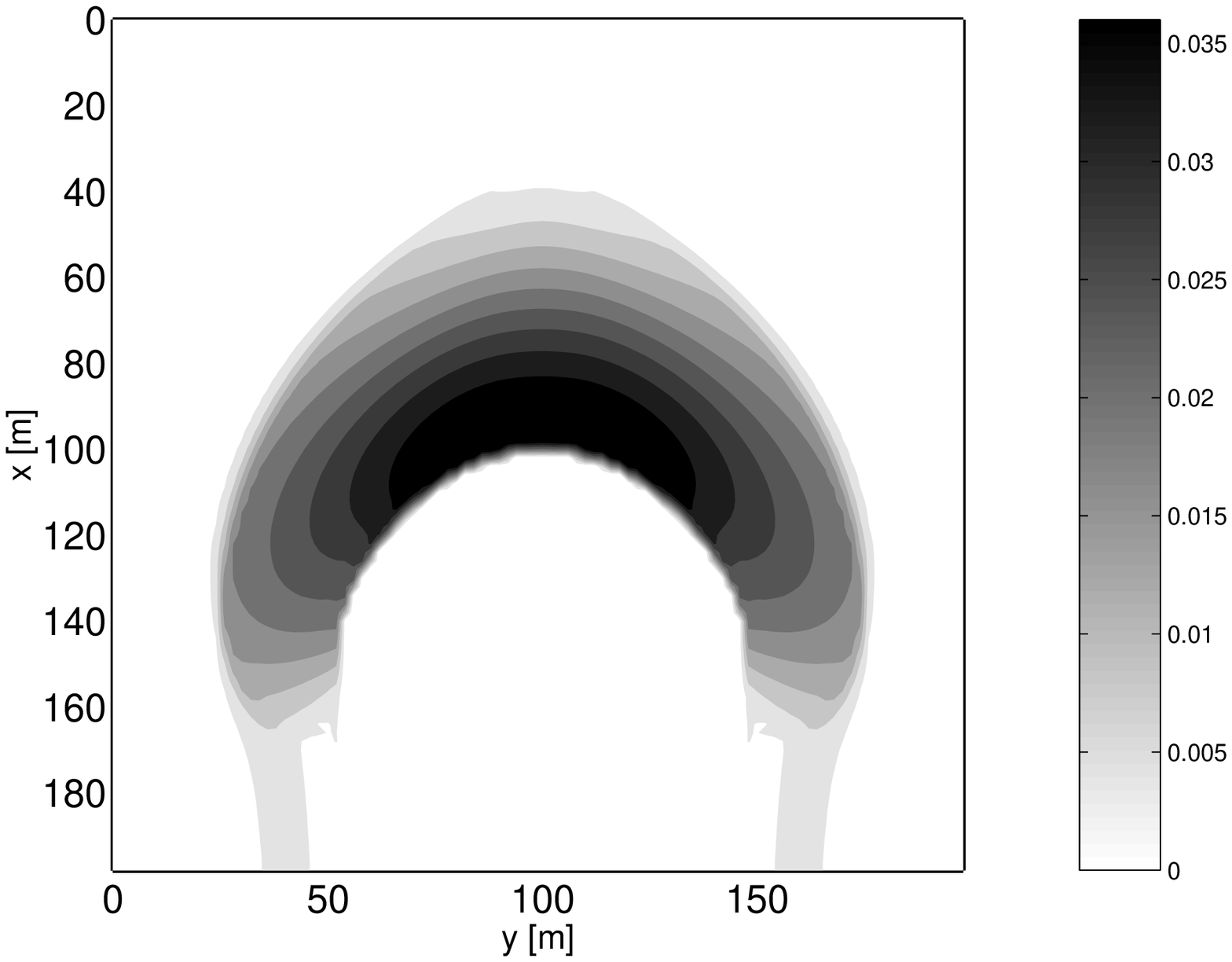}
    \includegraphics[width=0.45\textwidth]{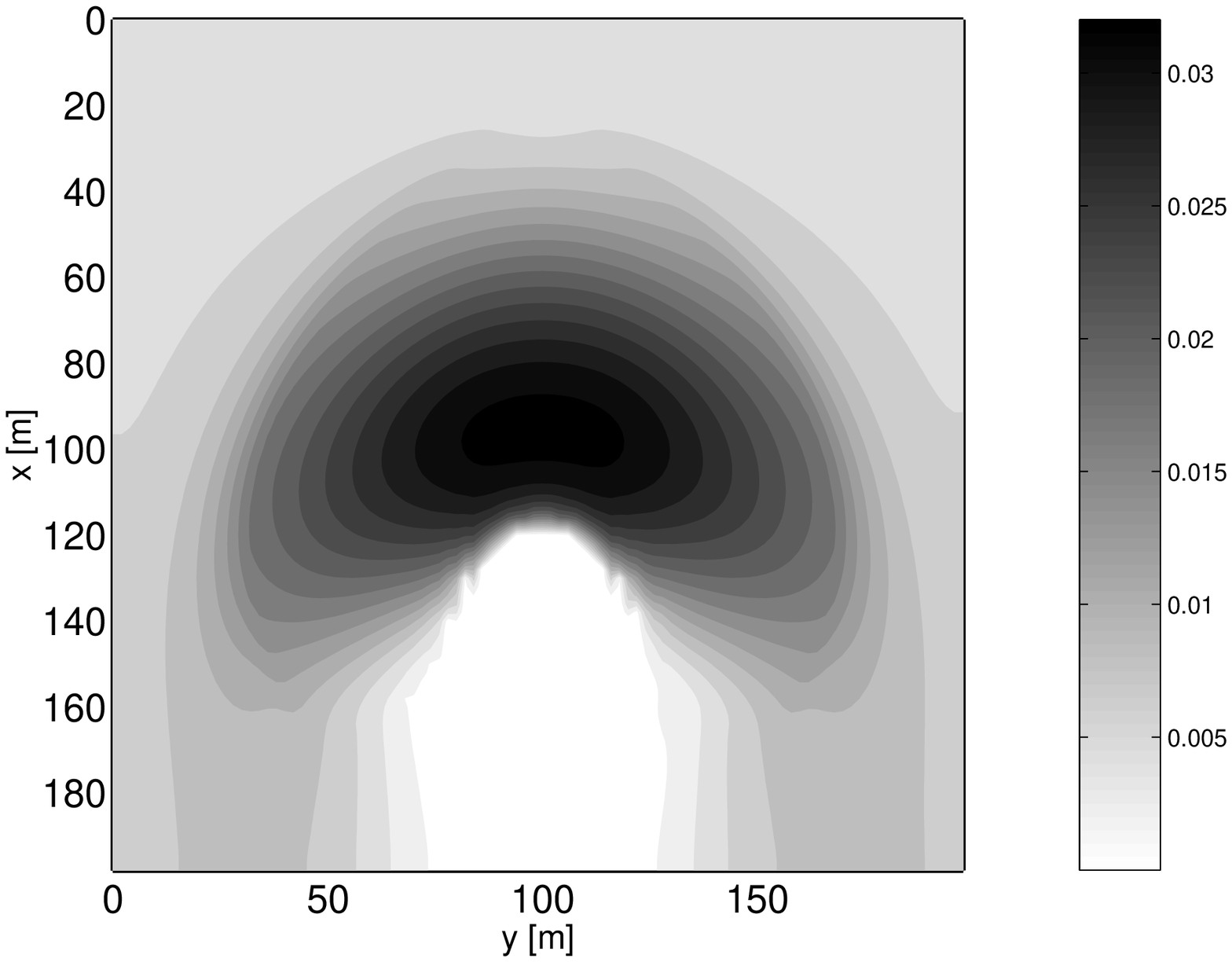}
\caption{Comparing the sand density $\rho$ of the saltation layer of two barchan 
         dunes of equal volume in steady state calculated from the same initial surface for 
         a diffusion constant of $C_{diff} = 0.0$ (top) and $C_{diff} = 4.0$ m$^2$s$^{-1}$ 
         (bottom). The diffusion term smoothes the sand density in the saltation layer.
         The shear velocity is $u_*=0.5$ms$^{-1}$. }
    \label{fig_barch_diffrho}
\end{figure}
%

\section{The effect of diffusion}
\label{sec:barch_effect_diff}
The dune model including the lateral component of the shear stress revealed a characteristic
shape with a rather sharp edge on the windward side which is not found for small barchan 
dunes for example in Morocco. To better understand this effect we
simulate barchan dunes of about 12 meters height having the same
volume  with different diffusion 
constants $C_{diff}$. For larger dunes and therefore larger 
length scales
the effect of diffusion transport will be weaker and the final state calculated by the 
dune model without diffusion should match real dune shapes. 

The results in this section can be discussed
only qualitatively due to the fact that experimentally the diffusion constant is unknown.
Figures~\ref{fig_barch_diffh} and 
\ref{fig_barch_diffrho} show the contour lines of the height profile and the sand density
in the saltation layer, respectively. Very high diffusion constants
like the one of the bottom part of
these figures show that the sharp edge in the shape of the windward side is 
smoothened. The sand density in the saltation layer  is flattened out as can be expected
from diffusion processes. Behind the brink always  the sand density drops to zero due to the 
vanishing shear stress. Diffusion changes the contour lines
of the shape  dramatically. An estimation of the diffusion 
constant and precise measurements of the height profiles of small and large barchans could 
give more insight into the influence of diffusion on the morphology of barchan dunes in the 
future.

\begin{figure}[htb]
  \center
    \includegraphics[width=0.45\textwidth]{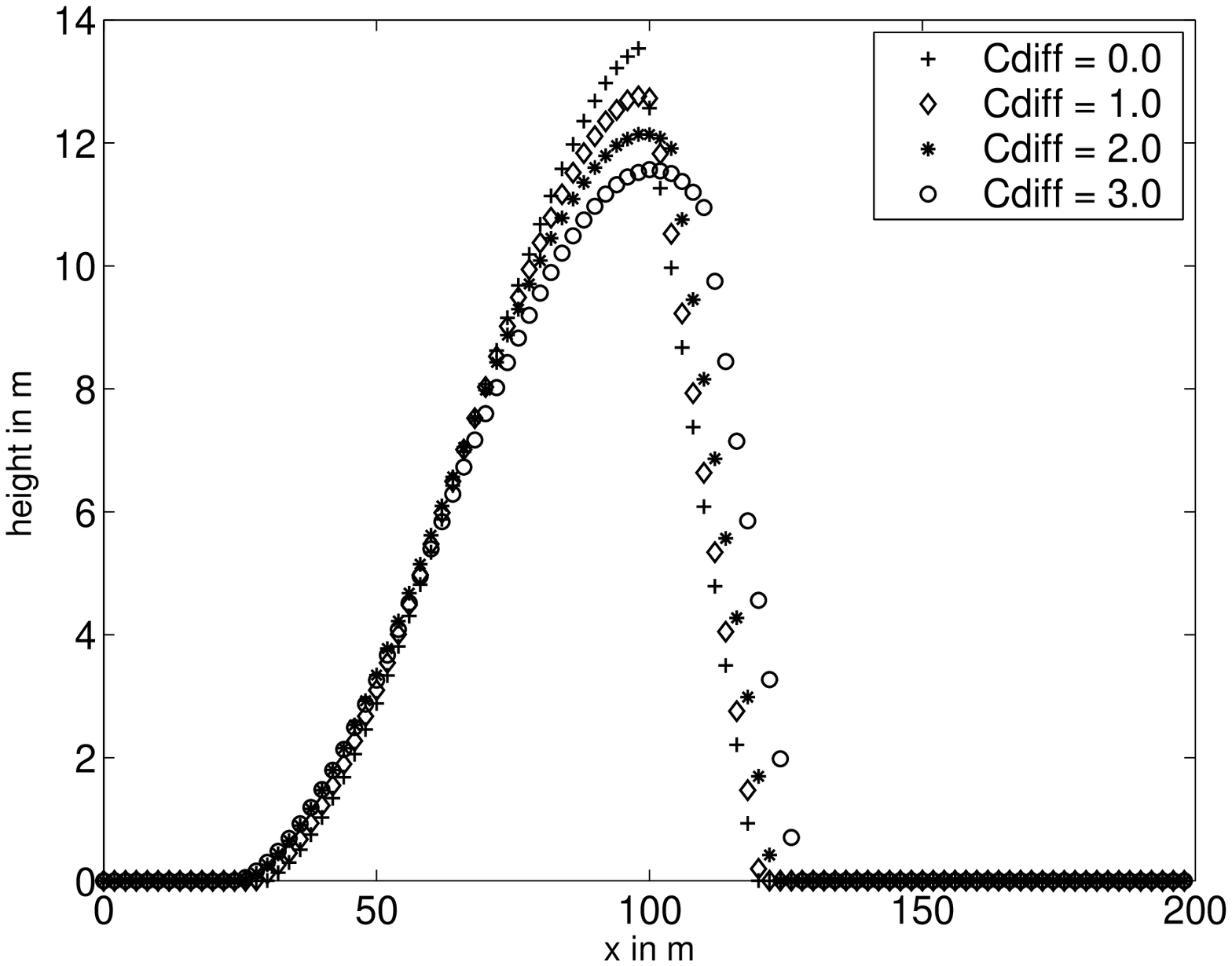}
\caption{The longitudinal cut through the dune for different diffusion constants $C_{diff}$.
         The length of the slip face  and the slope at the brink decrease with increasing
         diffusion constant. The shear velocity is $u_*=0.5$ms$^{-1}$. }
    \label{fig_barch_diffcrx}
\end{figure}

\begin{figure}[htb]
  \center
    \includegraphics[width=0.45\textwidth]{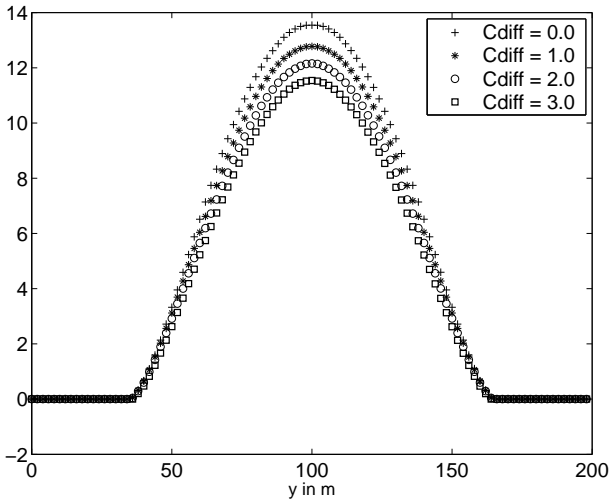}
\caption{The transversal cut through the windward side of the dune for different diffusion 
         constants $C_{diff}$.
         The height of the dune decreases with increasing diffusion constant. 
         The shear velocity is $u_*=0.5$ms$^{-1}$.}
    \label{fig_barch_diffcry}
\end{figure}

Figures~\ref{fig_barch_diffcrx} and \ref{fig_barch_diffcry} depict a
longitudinal and a  
transversal profile of a barchan dune in steady state for different diffusion constants. The height and the
slope at the brink decrease for a stronger diffusion. The length of the longitudinal profile
increases for a larger diffusion constant whereas  the width stays constant. The knowledge 
of $C_{diff}$ would also help to determine more realistic  values for $z_0$ and $L$ (Equation~(\ref{eq:tau_x}))
by fitting them to simulation results from a model with
diffusion. Consistency can be checked by comparing with  higher barchans where diffusion 
is weaker. But the computional costs and the missing measurements for large
barchans (there do not exist many large barchans) make the realization
still difficult at this point.

%

\section{Barchanoids, between barchan and transverse dunes}
\label{sec:barch_barchanoids}

If an area does not contain sufficient sand to form transverse dunes
but too much sand is available to keep the barchans isolated from each other 
the barchan dunes connect and
interesting hybrid forms appear, the so called barchanoids. The barchan dunes can be 
connected longitudinally and laterally. 

\begin{figure}[htb]
  \center
    \includegraphics[width=0.45\textwidth]{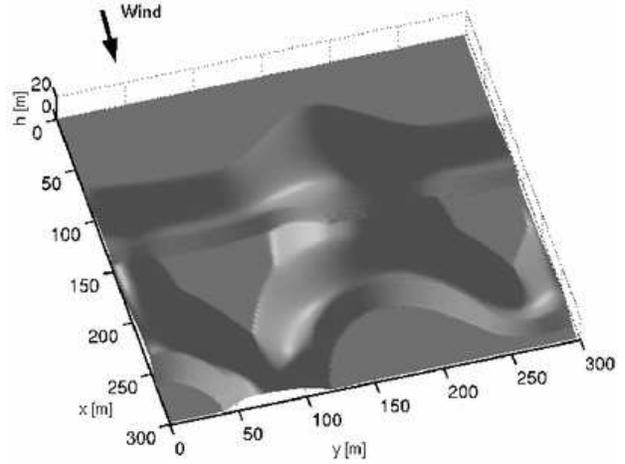}
\caption{A simulation with quasi-periodic boundary conditions. There is too much sand
         to build isolated barchans. The barchans are connected  in the longitudinal and 
         lateral direction. The shear velocity is $u_*=0.5$ms$^{-1}$}
    \label{fig_barch_barch1}
\end{figure}

\begin{figure}[htb]
  \center
    \includegraphics[width=0.45\textwidth]{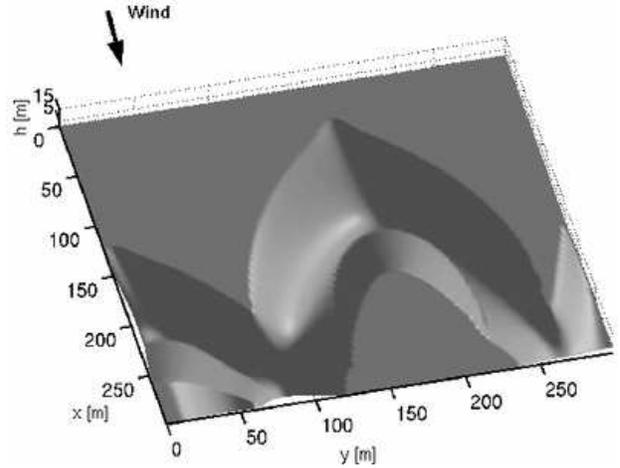}
\caption{Later state of the same simulation with quasi-periodic boundary conditions. Now the 
         barchans are connected only in the lateral direction. The shear velocity is 
         $u_*=0.5$ms$^{-1}$}
    \label{fig_barch_barch2}
\end{figure}

\begin{figure}[htb]
  \center
    \includegraphics[width=0.45\textwidth]{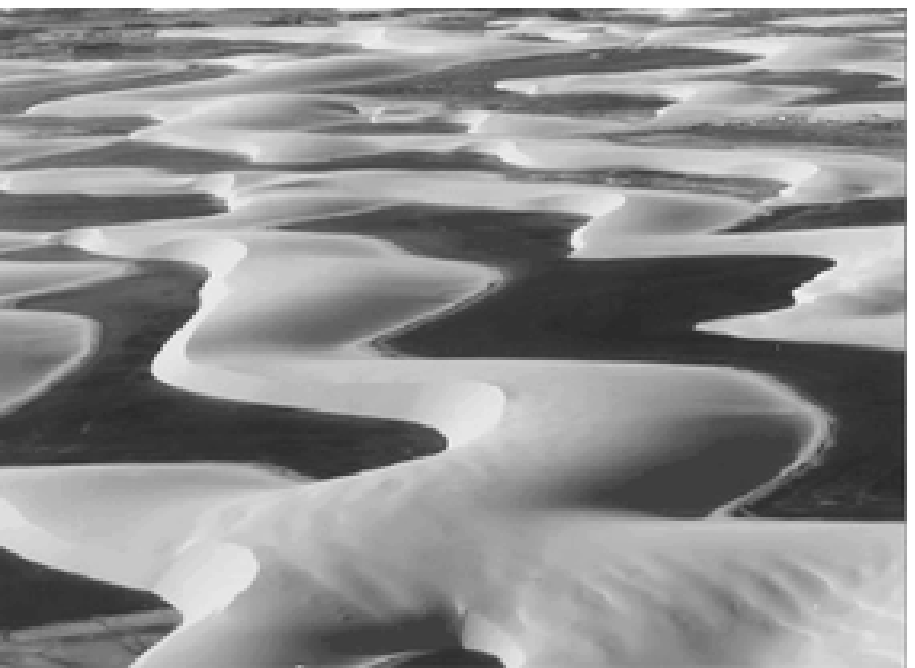}
\caption{Photo of the dune field of Len{\c c}ois Maranhenses, Brazil. In the rain season the 
inter-dune space is sometimes filled with lagoons. Barchan dunes are connected in longitudinal 
and lateral direction.}
    \label{fig_barch_barchphoto}
\end{figure}

Here only qualitative results will be presented. The simulation with the dune model 
is performed 
with a nearly flat initial surface (some small  Gaussian hills
randomly placed on a plane) of an averaged sand height of three meters. The boundary 
conditions are quasi periodic in wind direction and open in the lateral direction. The 
volume of sand in the simulation is held constant. Figures~\ref{fig_barch_barch1} and 
\ref{fig_barch_barch2} show two states of the simulation. In the first figure the barchans
are connected  longitudinally and laterally. The barchans are growing
so that the sand finally  
accumulates in a single row of larger dunes which are connected laterally. Similar barchanoids
can be found for example in the dune field of Len{\c c}ois Maranhenses in Brazil where 
on top in the rain season the
dunes can be separated by lagoons filled with rainwater (Figure~\ref{fig_barch_barchphoto}).
In this dune field the barchan dunes are connected in both directions. 
%

\section{Conclusions}
\label{sec:barch_concl}

The scaling laws revealed approximately linear relationships between height, width and length 
of the dunes. This is not valid for small dunes due to the influence of the saturation
length. Bagnold's law was modified relating the dune velocity to the length of the dunes.
This fit was surprisingly good and there seems to be a negligible influence of the 
saturation length on dune velocities. The transversal and the longitudinal cuts showed that
a parabola fits less well to numerical and real data than a $\sin^2$--function.
The shape invariance was verified and a characteristic sharp edge of the shape at the
windward side was observed in the calculations. 
The model was applied to the same dune
for different diffusion constants. The diffusion constant strongly influences the 
form of the final shape. Finally it was shown that barchanoids can 
 be obtained from longitudinally and laterally connected barchans.

\section{ Acknowledgement}

 H.J. Herrmann was partially funded by the Max-Planck-price.

\bibliographystyle{plain}
\bibliography{dune,unpublished}
\end{document}